\documentstyle[11pt]{article}
\newcommand{\be}{\begin{equation}}
\newcommand{\ee}{\end{equation}}
\newcommand{\bea}{\begin{eqnarray}}
\newcommand{\eea}{\end{eqnarray}}
\newcommand{\ba}{\begin{array}}
\newcommand{\ea}{\end{array}}
\newcommand{\nn}{\nonumber \\}

\newcommand{\NP}{Nucl.Phys.}

\newcommand{\PRL}{Phys.Rev.Lett.}

\begin{document}
\begin{titlepage}
\makebox[13cm][r]{IFUSP-P/1025}
\vspace{.5cm}
\begin{center}
\LARGE
{\sc TOPOLOGICAL TWO DIMENSIONAL DILATON SUPERGRAVITY}

\vspace{.5cm}
\large
V.O.Rivelles \footnote{Partially supported by CNPq}

{\it Instituto de F\'\i sica, Universidade de S\~ao
Paulo \\ C.Postal 20516, 01498, S.Paulo, SP, Brasil}

\vspace{.3cm}
December, 1992
\end{center}

\vspace{1cm}
\begin{abstract}
We present a topological version of two dimensional dilaton supergravity.
It is obtained by gauging  an extension of the super-Poincar\'e algebra
in two space-time dimensions. This algebra is obtained by an
unconventional contraction of the super de Sitter algebra. Besides the
generators of the super de Sitter algebra it has
one more fermionic generator and two more bosonic generators one of them being
a central charge. The gauging of this algebra is performed in the usual way.
Unlike some proposals for a dilaton supergravity theory
we obtain a model which is non-local in the gravitino field.
\end{abstract}
\end{titlepage}
\newpage
\large
Two dimensional gravity theories have been studied with the purpose of gaining
conceptual and technical insights in order to handle the more difficult problem
of treating four dimensional gravity \cite{JackTeit}.
Another point of interest is that effective string theories in four dimensions
show that Einstein gravity should be corrected by coupling to a dilaton field
\cite{Callan}. In
particular the process of black-hole evaporation \cite{Hawk} should then be
studied in
this new framework \cite{Evan}. In this case the problem can be reduced to a
two-dimensional dilaton gravity model coupled to topological matter. When
we consider
only the graviton and the dilaton fields the resulting theory is a topological
gravity theory\cite{Verl}.

The introduction of supersymmetry in dilaton gravity theories was made in a
non topological way. It was
motivated in order to have a positive energy theorem for the two dimensional
case \cite{Park} and also for the study of two-dimensional black holes
\cite{Nojiri}. On the other hand topological supergravity theories have been
obtained.  The
construction of $N=1$ \cite{Mont} and $N=2$ \cite{Li2} topological
supergravities have
been made by gauging the groups $OSp(2 \mid 1)$ and $OSp( 2 \mid 2)$,
respectively. By a contraction of the gauge group the corresponding
super-Poincar\'e theories are then obtained.

As stated above the dilaton gravity theory can have a gauge formulation leading
to a topological theory \cite{Verl}.
When the gauge
algebra is the Poincar\'e algebra the fields have unusual transformation
properties under Poincar\'e gauge transformations \cite{Verlpre}. This can be
overcome if we
take a central extension of Poincar\'e algebra as the gauge algebra
\cite{Cange}. This
central extension consists in changing the usual commutator of the translation
generators of the Poincar\'e algebra  to $[P_a, P_b] =
\epsilon _{ab} Z$ where $Z$ belongs to the center of the algebra. This new
algebra can be obtained by an unconventional contraction of the de Sitter
algebra $SO(2,1)$
where the Lorentz generator $J$ is replaced by $J + Z/ \lambda$, where
$\lambda$ is the
cosmological constant, and the limit
$\lambda \rightarrow 0$ is taken \cite{Cange}.

In this paper we generalize this unconventional contraction to the super de
Sitter $OSp(2 \mid 1)$ algebra to obtain the supersymmetric version of the
central extension of
the Poincar\'e algebra. We will show that besides the supersymmetry generator
$Q_\alpha$ it is necessary another fermionic generator $U_\alpha$ which does
not generate any supersymmetry. It is also necessary the introduction of a
scalar generator $K$ besides the central charge of the non supersymmetric case.
We find the quadratic Casimir operator starting from that of the $SO(2,1)$
algebra showing the existence of a non-degenerated metric. We then gauge this
new algebra and build up the corresponding topological supergravity theory.

In two dimensions the commutation relations
\bea
\label{desitter}
{}[P_a, P_b] &=& \lambda \epsilon_{ab} J \nn
{}[J, P_a] &=& {\epsilon_a}^b P_b
\eea
define the $SO(2,1)$ and $SO(1,2)$ algebras for $\lambda > 0$ and $\lambda < 0$
respectively. This is independent of the conventions for the flat Minkowski
metric $h_{ab}$ and the antisymmetric tensor $\epsilon_{ab}$. For definiteness
(which is needed in the supersymmetric case) we will adopt the conventions
$h_{ab} = diag(-1, +1)$ and $\epsilon^{01} = 1$. The supersymmetric extension
of the algebra (\ref{desitter}) with $\lambda > 0$ and with $N = 1$
supersymmetry is \cite{Mont}
\bea
\label{sdesitter}
{}[P_a, P_b] &=& \lambda \epsilon_{ab} J \nn
{}[J, P_a] &=& {\epsilon_a}^b P_b \nn
{}[J, Q_\alpha ] &=& - \frac{1}{2} (\gamma_2 Q)_\alpha \nn
{}[P_a, Q_\alpha] &=& \frac{1}{2} \lambda^\frac{1}{2} (\gamma_a Q)_\alpha \nn
\{ Q_\alpha, Q_\beta \} &=& (\gamma^a C)_{\alpha \beta} P_a -
\lambda^\frac{1}{2} (\gamma_2 C)_{\alpha \beta} J
\eea
This is the algebra $OSp(2 \mid 1)$ written in a Lorentz covariant form.
The conventions for the Dirac
matrices are the following $\gamma^0 = i \sigma_2, \gamma^1 = \sigma_1,
\gamma_2 = \sigma_3$ and $C = i \sigma_2$ where $\sigma_i$ are the Pauli
matrices. For $\lambda < 0$ the chiral components of $Q_\alpha$ acquire
different signs in the commutator with $P_a$ and $J$ and they give rise to the
algebra $OSp(1 \mid 2)$.

The quadratic Casimir operator of the algebra (\ref{sdesitter}) is
\be
\label{casimir}
C = P^2 + \lambda J^2 + \frac{1}{2} \lambda^\frac{1}{2} Q_\alpha C^{\alpha
\beta} Q_\beta
\ee
which gives rise to the following non-degenerated metric on the algebra
$<P_a, P_b> =
h_{ab}, <J, J> = \lambda$ and $<Q_\alpha, Q_\beta> = \frac{1}{2}
\lambda^\frac{1}{2} C_{\alpha \beta}$.

We will now perform in a systematic way the unconventional contraction
\cite{Cange} which leads to the central extension of the algebra
(\ref{desitter}) in order to extended it to the supersymmetric case later on.
Due to the well known ambiguity of the angular momentum in two dimensions we
first introduce a new generator $Z$ through the replacement
\be
\label{repl1}
J \rightarrow J + Z/\lambda
\ee
To get all commutation relations involving this new
generator $Z$ we will perform the replacement (\ref{repl1}) in
(\ref{desitter}) and equate the terms with the same power of $\lambda$
which become singular  when we  take the
limit $\lambda \rightarrow 0$.
{}From the commutation relation $[J,P_a]$ we find $[J, P_a] =
{\epsilon_a}^b P_b$ and $[Z, P_a] = 0$ and from
the commutation relation $[P_a, P_b]$ we find $[P_a, P_b] =
\epsilon_{ab} Z$. Taking into account the ambiguity in the angular momentum
we impose that the old commutation relation $[J, J] = 0$ should be replaced
by $[J, J + Z/\lambda] = 0$ and we find that $[J,J] = [J, Z] = 0$. We then
conclude that $Z$ belongs to the center of the
algebra and we get the central extension of the Poincar\'e algebra
\cite{Cange}
\bea
\label{central}
{}[P_a, P_b] &=& \epsilon_{ab} Z \nn
{}[J, P_a] &=& {\epsilon_a}^b P_b \nn
{}[P_a, Z] &=& [J, Z] = 0
\eea

We will now apply the same procedure to the supersymmetric case
(\ref{sdesitter}).
In the anticommutation relation $\{Q_\alpha, Q_\beta\}$ there will
appear a term proportional to $\lambda^{-\frac{1}{2}} Z$ on the right hand
side of
it. This means that a replacement of $Q_\alpha$ involving powers of
$\lambda^{-\frac{1}{2}}$
is needed which implies the introduction of a new fermionic generator. Since
powers of $\lambda^{-\frac{1}{2}}$ will appear in the commutators they must
also be allowed in
(\ref{repl1}) which means the introduction of a new bosonic generator. So we
start with the following replacements
\bea
\label{repl2}
J &\rightarrow& J + \lambda^{-1} Z +  \lambda^{-\frac{1}{2}} K \nn
Q_\alpha &\rightarrow& Q_\alpha + \lambda^{-\frac{1}{2}}  U_\alpha
\eea
where $K$ is the new bosonic generator and $U_\alpha$ is the new fermionic
generator. The replacements (\ref{repl2}) just reflect the ambiguity of the
angular momentum extended to the supersymmetry generators.
It is not difficult to find out
the non-vanishing commutation relations according to the rules stated above.
\bea
\label{scentral}
{}[P_a, P_b] &=& \epsilon_{ab} Z \nn
{}[J, P_a] &=& {\epsilon_a}^b P_b \nn
{}[P_a, Q_\alpha] &=& \frac{1}{2} ( \gamma_a U)_\alpha \nn
{}[J, Q_\alpha] &=& - \frac{1}{2} ( \gamma_2 Q)_\alpha \nn
{}[J, U_\alpha] &=& - \frac{1}{2} ( \gamma_2 U)_\alpha \nn
{}[K, Q_\alpha] &=& - \frac{1}{2} ( \gamma_2 U )_\alpha \nn
\{Q_\alpha, Q_\beta \} &=& ( \gamma^a C )_{\alpha \beta} P_a - ( \gamma_2 C
)_{\alpha \beta} K \nn
\{Q_\alpha, U_\beta \} &=& - ( \gamma_2 C )_{\alpha \beta} Z
\eea
We can check that all the Jacobi identities are
satisfied.  Then the generator $U_\alpha$ transform as a spinor under
a Lorentz transformation but does not
generate any supersymmetry since its anticommutator vanishes.
$K$ behaves like a
momentum generator in a third direction regarding the fermionic sector,
while in the bosonic
sector it behaves like a central charge. $Z$ remains a central charge.
We can set $Z = K =
U_\alpha = 0$ consistently and we then obtain the $N = 1$ super-Poincar\'e
algebra.
Therefore the algebra
(\ref{scentral}) is a non-trivial extension of the super-Poincar\'e algebra.

We can find the quadratic Casimir operator by making the replacement
(\ref{repl2}) in (\ref{casimir}). We find besides the regular term given
below terms
proportional to $\lambda^{-1}$ and $\lambda^{-\frac{1}{2}}$. These terms
are also
invariants of the algebra but they give rise to a degenerated metric on the
algebra. The regular term gives the quadratic Casimir operator
\be
\label{newcasimir}
C = P^2 + K^2 + JZ + ZJ + \frac{1}{2} C^{\alpha \beta} ( Q_\alpha U_\beta  +
U_\beta Q_\alpha )
\ee
from which we can read off the metric $<P_a, P_b> = h_{ab}, <J, Z> = 1$,
$<K, K> = 1, <Q_\alpha, U_\beta> = \frac{1}{2} C_{\alpha \beta}$ which is
non-degenerated.

The topological gauge theory associated to this algebra follows in the usual
way \cite{Witten}. The one-form gauge potential $A$ has the following
expansion in terms of the generators of the algebra
\be
\label{pot}
A = e^a P_a + w J + v K + \psi^\alpha Q_\alpha + \xi^\alpha U_\alpha + A Z
\ee
The two-form field strength $F = dA + A^2$ can be used to write the action
$S = \int Tr ( \eta F )$ where $\eta$ is in the coadjoint representation. We
then get
\bea
\label{action}
S &=& \int [ \eta_a F^a(P) + \eta F(J) + \eta^\prime F(K) + \nn
&+& \eta^{\prime \prime} F(Z) + \chi_\alpha F^\alpha (Q) + \chi^\prime_\alpha
F^\alpha (U) ]
\eea
where the field strengths are
\be
\label{FA}
F^a(P) = d e^a + w e^b {\epsilon_b}^a - \frac{1}{2} \psi \gamma^a \psi
\ee
\be
\label{FJ}
F(J)   = dw
\ee
\be
\label{FK}
F(K)   = dv + \frac{1}{2} \psi \gamma_2 \psi
\ee
\be
\label{FZ}
F(Z)   = dA + \frac{1}{2} e^a e^b \epsilon_{ab} + \psi \gamma_2 \xi
\ee
\be
\label{FQ}
F^\alpha(Q) = D\psi^\alpha \equiv d\psi^\alpha - \frac{1}{2} w (\psi
\gamma_2)^\alpha
\ee
\be
\label{FU}
F^\alpha(U) = D\xi^\alpha + \frac{1}{2} e^a (\psi \gamma_a)^\alpha -
\frac{1}{2} v (\psi \gamma_2)^\alpha
\ee
The field equations obtained from the variation of $\eta$ are then
$F = 0$ and those obtained from the variation of the gauge fields are
\be
\label{etaa}
D\eta_a + \epsilon_{ab} e^b \eta^{\prime \prime} - \frac{1}{2} \psi \gamma_a
\chi^\prime = 0
\ee
\be
\label{eta}
d\eta + e^a {\epsilon_a}^b \eta_b + \frac{1}{2} \psi \gamma_2 \chi +
\frac{1}{2} \xi \gamma_2 \chi^\prime = 0
\ee
\be
\label{etap}
d\eta^\prime + \frac{1}{2} \psi \gamma_2 \chi^\prime = 0
\ee
\be
\label{etapp}
d\eta^{\prime \prime} = 0
\ee
\be
\label{chi}
D\chi + \eta_a \gamma^a \psi - \eta^\prime \gamma_2
\psi - \eta^{\prime \prime} \gamma_2 \xi - \frac{1}{2} e^a
\gamma_a \chi^\prime + \frac{1}{2} v \gamma_2 \chi^\prime = 0
\ee
\be
\label{chip}
D\chi^\prime - \eta^{\prime \prime} \gamma_2 \psi = 0
\ee

Equation (\ref{FA}) allows us to solve for the spin connection
\be
\label{conn}
w = - (det e)^{-1} e^a \epsilon^{\mu \nu} ( \partial_\mu e_\nu^b  h_{ab} -
\frac{1}{2} \psi_\mu \gamma_a \psi_\nu )
\ee
Then equation (\ref{FJ}) implies that $R = 0$, where $R$ is curvature
scalar, while (\ref{FQ}) implies the vanishing of the gravitino field strength.
{}From (\ref{etapp}) we obtain  $\eta^{\prime \prime} = \lambda =
constant$ which plays the role of the cosmological constant as in the
bosonic case \cite{Cange}. The remaining fields may have a geometric
interpretation when coupled to matter (like in the bosonic case
\cite{Cangelast}). In order to get the corresponding supergravity
theory we solve the equations for $v, \eta_a, \eta^\prime$ and $\xi$ obtaining
\bea
\label{eqnsolved}
v &=& \frac{1}{2 \eta^{\prime \prime}} \chi^\prime \psi \nn
\eta_a &=& - {\epsilon_a}^b  e_b^\mu ( \partial_\mu \eta + \frac{1}{2} \psi_\mu
\gamma_2 \chi + \frac{1}{2} \xi_\mu \gamma_2 \chi^\prime ) \nn
\eta^\prime &=& \frac{1}{4 \eta^{\prime \prime}} \chi^\prime \chi^\prime \nn
\xi &=& \frac{1}{2 \eta^{\prime \prime}} e^a \chi^\prime \gamma_a \gamma_2 +
\frac{1}{8 ( \eta^{\prime \prime} )^2} (\chi^\prime \chi^\prime) \psi
\eea
We then find that the action (\ref{action}) reduces to
\be
\label{sugraaction}
S = \int [ \eta dw + \chi_\alpha D \psi^\alpha + \frac{1}{2} \eta^{\prime
\prime} e^a e^b \epsilon_{ab} - \frac{1}{2} e^a \psi \gamma_a \chi^\prime +
\frac{1}{8} \eta^{\prime \prime} \chi^\prime \chi^\prime  \psi \gamma_2 \psi ]
\ee
The supersymmetry transformations can be obtained from the transformations
generated by $Q$ and using the solutions
(\ref{eqnsolved}). The resulting supersymmetry transformations which leave the
action (\ref{sugraaction}) invariant are
\bea
\label{sugra}
\delta e^a = \epsilon \gamma^a \psi, & & \delta \eta = -\frac{1}{2} \epsilon
\gamma_2 \chi \nn
\delta \psi = D \epsilon, & & \delta \chi = - \eta_a \gamma^a \epsilon \nn
\delta \chi^\prime = \eta^{\prime \prime} \gamma_2 \epsilon & &
\eea
where the equation relating $\chi^\prime$ and $\psi$ (\ref{chip}) has been
used. If we eliminate $\chi^\prime$ from the action by using the field equation
(\ref{chip}) then the resulting action would be non local.  We then have a
non local version of $N=1$
topological supergravity theory. The non-locality appears only in the fermionic
sector and it may give rise to some new class of induced supergravity theory.
Other aspects, like the relation to matrix models and the presence (or absence)
of black holes remains to be investigated.

Notice however that there exists a local
version for $N=1$ dilaton supergravity \cite{Park,Nojiri}. It is given by
the action (after some field redefinitions and elimination of the auxiliary
fields)
\bea
\label{localaction}
S &=& \int [ \eta dw + \chi_\alpha D \psi^\alpha + \eta^{\prime \prime} e^a e^b
\epsilon_{ab} + \nn
&+& (\eta^{\prime \prime})^2 ( \eta^{\frac{1}{2}} \psi \gamma_2 \psi
+ \frac{1}{2} \eta^{-\frac{1}{2}} e^a \psi \gamma_a \psi + \frac{1}{16} \chi
\chi e^a e^b \epsilon_{ab} ) ]
\eea
This action is invariant under the following supersymmetry transformations:
\bea
\label{localtransf}
\delta e^a = \epsilon \gamma^a \psi, & & \delta \eta = - \frac{1}{2}
\epsilon \chi \nn
\delta \psi = D \epsilon - \frac{1}{2} (\eta^{\prime \prime})^{\frac{1}{2}}
e^a \epsilon \gamma_a, & & \delta \chi = - \eta_a \gamma^a \epsilon + 2
(\eta^{\prime \prime})^\frac{1}{2} \eta^{\frac{1}{2}} \gamma_2 \epsilon
\eea
However, as it is easily seen,
the field equations do not imply the zero curvature condition neither the
vanishing of the gravitino field strength. So, although it is a possible
extension of the dilaton gravity theory it does not share the topological
properties presented by (\ref{sugraaction}).

\pagebreak


\begin{thebibliography}{99}

\bibitem{JackTeit}R.Jackiw, in {\it Quantum Theory of Gravity}, ed.
S.Christensen (Hilger, Bristol, 1984); C.Teitelboim, Phys.Lett. {\bf 126B}
(1983) 41, in {\it Quantum Theory of Gravity}, ed.
S.Christensen (Hilger, Bristol, 1984).

\bibitem{Callan}C.G.Callan, D.Friedan, E.J.Martinec and M.J.Perry, Nucl. Phys.
{\bf B262} (1985) 593.

\bibitem{Hawk}S.W.Hawking, Comm.Math.Phys. {\bf 43} (1975) 199.

\bibitem{Evan}C.G.Callan, S.B.Giddings and J.A.Harvey, Phys.Rev.D {\bf 45}
(1992) R1005.

\bibitem{Verl}E.Verlinde and H.Verlinde, Nucl.Phys. {\bf B348} (1991) 457.

\bibitem{Park}Y.Park and A.Strominger, {\it Supersymmetry and Positive Energy
in Classical and Quantum Two-Dimensional Dilaton Gravity}, preprint
UCSBTH-92-39 (1992).

\bibitem{Nojiri}S.Nojiri and I.Oda, {\it Dilatonic Supergravity in Two
Dimensions
and the Disappearance of Quantum Black Hole}, preprint NDA-FP-8/92, OCHA-PP-30
(1992).




\bibitem{Mont}D.Montano, K.Aoki and J.Sonnenschein, Phys.Lett. {\bf 247B}
(1990) 64.

\bibitem{Li2}K.Li, Nucl.Phys. {\bf B346} (1990) 329.

\bibitem{Verlpre}H.Verlinde, {\it Black Holes and Strings in Two Dimensions},
preprint PUPT-1303 (1991).

\bibitem{Cange}D.Cangemi and R.Jackiw, \PRL {\bf 69} (1992) 233.

\bibitem{Witten}E.Witten, \NP {\bf B340} (1990) 333.

\bibitem{Cangelast}D.Cangemi and R.Jackiw, {\it Geometric Gravitational Forces
on Particles Moving in a Line}, preprint CTP 2147 (1992).

\end{thebibliography}
\end{document}